\documentclass[useAMS,usenatbib]{mn2e}
\usepackage{graphicx}
\usepackage{amssymb}
\def\kk{}
\title[Hard TeV spectra of blazars]{Hard TeV spectra of blazars and the constraints to the IR
       intergalactic background}
\author[K.~Katarzy\'nski et al.]
       {K.~Katarzy\'nski,$^{1,2}$\thanks{E-mail:kat@astro.uni.torun.pl}
        G.~Ghisellini,$^{1}$
        F.~Tavecchio,$^{1}$
        J.~Gracia,$^{3}$
        L.~Maraschi,$^{1}$\\
$^{1}$Osservatorio Astronomico di Brera, via Bianchi 46, Merate and via Brera 28, Milano, Italy\\
$^{2}$Toru\'n Centre for Astronomy, Nicolaus Copernicus University, ul. Gagarina 11, PL-87100 Toru\'n, Poland\\
$^{3}$IASA, Dept. of Physics, Univ. of Athens, Panepistimiopolis, 15784 Zografos, Athens\\
}
\begin{document}

\date{Accepted 2006 January 31; Received 2005 November 16; in original form 2005 November 16}

\pagerange{\pageref{firstpage}--\pageref{lastpage}} \pubyear{2005}

\maketitle

\label{firstpage}

\begin{abstract}
Recent gamma--ray observations of the blazar 1ES~1101--232 (redshift
$z=0.186$) reveal that the unabsorbed TeV spectrum is hard, with
spectral index $\alpha \lesssim 0.5$ [$F(\nu) \propto
\nu^{-\alpha}$]. We show that simple one--zone synchrotron
self--Compton model can explain such hard spectra if we assume a power
law energy distribution of the emitting electrons with a relatively
high minimum energy.  In this case the intrinsic TeV spectrum can be
as hard as $F(\nu)\propto \nu^{1/3}$, while the predicted X--ray
spectrum can still be much softer. The observations of 1ES~1101--232
can therefore be reconciled with relatively high intensities of the
infrared background, even if some extreme background levels can indeed
be excluded. We show that the other TeV sources (Mrk~421, Mrk~501 \&
PKS 2155--304) can be interpreted in the same framework, with a
somewhat larger minimum energy.
\end{abstract}


\section{Introduction}

The intrinsic high-energy emission from extragalactic sources is
attenuated by the pair-production process (\citealt{Nishikov62},
\citealt{Gould66}, \citealt{Stecker92}). Photons in the TeV range
($\epsilon_{\gamma}$) interact with low--energy photons from the
extragalactic infrared background ($\epsilon_{\rm IR}$) and produce
electron--positron pairs ($\epsilon_{\rm IR} + \epsilon_{\gamma}
\rightarrow e^+e^-$). The total absorption depends on the history of
the reaction rate along the line--of--sight, and therefore, on the
distance travelled and on the local density of the low-energy target
photon field, both a function of redshift $z$. This process will
affect the high--energy TeV end of the observed 
blazar spectra. Since the optical depth depends strongly on
the incident TeV--photon energy the process leads to a steepening of the
observed spectrum with respect to the intrinsic (unabsorbed) one.

Thus, the observed spectra from distant blazars contain information on
both the IR background radiation field history and the intrinsic
properties of the source. From the viewpoint of blazar studies the
latter is just an undesirable extrinsic effect that must be corrected
for. In contrast, from the opposite perspective, it allows the
measurement of the IR background. In both cases these two independent
processes have to be disentangled by either modeling both
simultaneously (e.g. \citealt{DeJager02}, \citealt{Kneiske04}), or 
measuring one independently (see \citealt{Hauser01} for a review).

Unfortunately it proved very difficult to constrain the IR background
independently, and different groups tried to constrain models for the
IR background from TeV observations {\em assuming} that the blazar
spectrum is a well-behaved power--law (e.g. \citealt{Costamante03},
\citealt{Dwek05}). The best perspective for such studies is offered by
high--redshift blazars, since the absorption will be strong. Recently,
\citet{Aharonian06} presented a study along these lines, based on the
TeV spectra of high--redshift blazars recently measured by the
H.E.S.S. instrument.
They de--reddened the observed TeV data of 1ES~1101-232 with different
models of the IR backgrounds, and concluded that the resulting
intrinsic spectrum is {\it always} very hard.
The index of the intrinsic spectrum should be equal to $\alpha=0.5$ if
the level of the IR background is close to the absolute lower limit
obtained from direct integration of the light from resolved galaxies
\citep{Madau00}.  Larger IR background levels should correspond to
intrinsic TeV spectra harder than $\alpha=0.5$. Note that this
spectral index is already harder than the spectral index observed in
other nearby TeV sources and that the difference will be more
important for larger background levels. \citet{Aharonian06} also
commented on the theoretical difficulties to explain spectra harder
than $\alpha=0.5$, which would imply quite unexpected physics.  These
facts led \citet{Aharonian06} to prefer the solution of minimal IR
background.

In this paper we show instead that spectra harder than $\alpha=0.5$
can be obtained in ``normal" one--zone homogeneous synchrotron
self--Compton (SSC) models, as long as there is a deficit of low
energy electrons with respect to the extrapolation from higher
energies.  This can be achieved by assuming an emitting particle
distribution which is a double power law, with a flat slope ($<2$) of
the first, low energy portion, or assuming a simple power law, but
with a relatively high low energy cut--off.  In these cases the limits
on how hard a TeV spectrum can be is $\alpha=-1/3$, while well
accounting for a much softer X--ray spectrum.

We discuss this result with the aim of finding limits on the IR
background and also in the context of explaining the other TeV sources
(with a softer intrinsic spectrum) with the same model.

\section{SSC models}

The simplest possible scenario able to explain the X--ray and TeV
emission of a blazar assumes a homogeneous, spherical source filled
by tangled magnetic field and relativistic electrons. 

Since in TeV blazars we must have electrons of very large energies,
the scattering process occurs partly in the Klein--Nishina regime.
In this regime the scattering process
becomes inefficient, and for typical electron distributions the SSC
process can be well described by approximating the scattering cross
section with a step function, equal to the Thomson cross section for
incident photon energies $h \nu < m_{\rm e}c^2/\gamma$, and zero
otherwise\footnote{ The models shown in all figures and all the
numerical results made use the full Klein--Nishina cross section; this
approximation is mentioned here for illustrative purposes only.  }.
This means that electrons of increasingly large energies scatter an
increasingly smaller fraction of synchrotron seed photons, steepening
the resulting SSC spectrum with respect to the ``canonical" value of
$\alpha=(n-1)/2$.  Note that this steepening introduced by the
Klein--Nishina effects is virtually inescapable, since the scattering
occurs completely in the Thomson limit only for extremely large values
of the beaming factor: $\delta>1000$ (see
e.g. \citealt{Katarzynski05a}). In the next section we will find what is
the high energy spectral index corresponding to a given particle
energy distribution, assuming ``usual" parameters for our TeV source,
namely: source radius $R=10^{15\to 16}$ [cm], magnetic field intensity
$B=1\to0.01$ [G], Doppler factor $\delta=10\to50$ and approximating the 
particle energy distribution by a broken power law with the index $n_1$ 
[$N(\gamma) = K \gamma^{-n_1}$] below the break energy ($\gamma_{\rm brk}$) 
{\kk
and $n_2$ above the break [$N(\gamma) = K \gamma_{\rm brk}^{n_2-n_1} 
\gamma^{-n_2}$], where $K$ is the density of the particles with $\gamma=1$. 
If the minimal $\gamma$ is greater than unity ($\gamma_{\rm min} > 1$)
the density of the particles with minimum energy ($K'$) is 
given by $K' = K \gamma_{\rm min}^{-n_1}$.
}

\begin{figure}
\resizebox{\hsize}{!}{\includegraphics{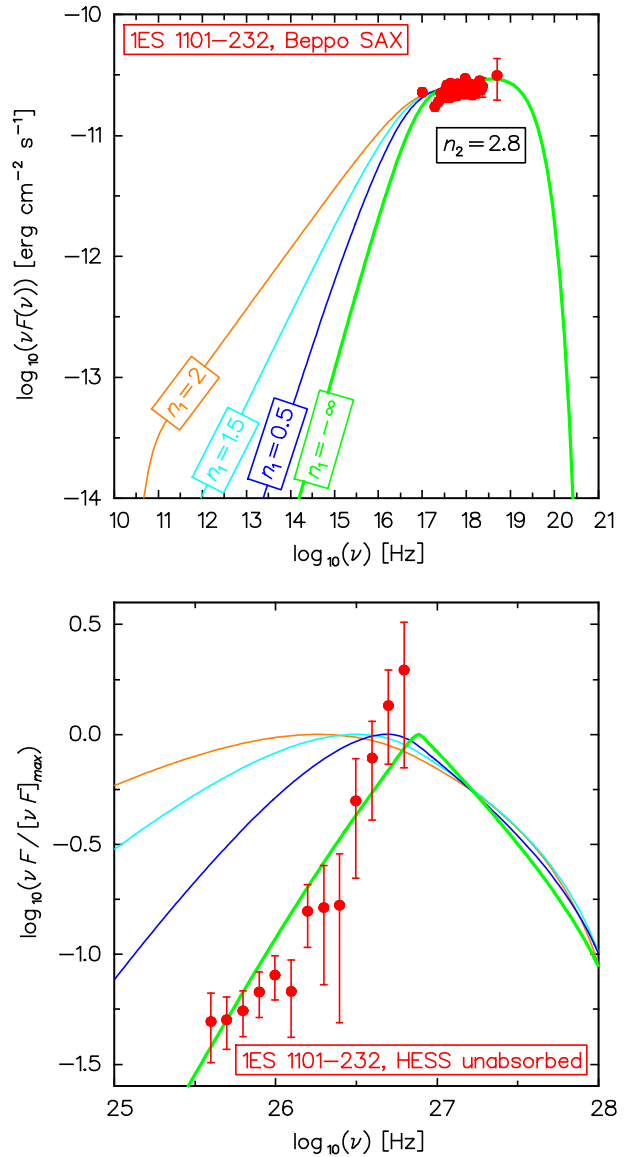}}
\caption{Synchrotron emission (upper panel) and self--Compton
         radiation (lower panel) for different values of $n_1$. The
         green spectra, generated by the truncated single power law
         particle energy spectrum ($n_1=-\infty$), provide the best
         fit for observations of 1ES~1101--232 made by $Beppo$SAX
         (Wolter et al. 2002) and H.E.S.S. (Aharonian et
         al. 2006). Note that the spectra on the lower panel have been
         normalized to the peak level and the H.E.S.S. observations
         has been unabsorbed according to the model proposed by
         Kneiske et al. (2004).}
\label{fig_n1t}          
\end{figure}

\subsection{Flat electron distributions}

A flat slope of the electron distribution ($n_1 < 2$) may help to
explain the intrinsic TeV spectrum of 1ES 1101--232.  Assuming the
lowest possible absorption level, \citet{Aharonian06} have estimated
the intrinsic spectral index $\alpha \simeq 0.5$ for this particular
source. We verified that it is roughly possible to reproduce such a
spectrum if we assume that $n_1 \simeq 1.5$ (note that in the
Thomson limit $\alpha=0.5$ would correspond to $n_1=2$.).  In this
case, however, we do not obtain a power law intrinsic TeV spectrum.
The spectrum is curved, with $\alpha$ changing from $\alpha=0.25$ in
the GeV range to $\alpha=1$ at the peak of the emission at a few TeV.

Since $n_1 \simeq 1.5$ can explain the spectrum derived for the
minimal IR background case, we then ask if a smaller value of $n_1$,
producing an harder spectrum, can allow for a larger level of the IR
background.  In Fig. \ref{fig_n1t} we show different synchrotron and
inverse--Compton spectra for different $n_1$ indices.  The spectral
index of the first part of the thin synchrotron emission ($\alpha_1$)
depends directly on $n_1$ [$\alpha_1=(n_1-1)/2$] down to the limiting
value $n_1=1/3$.  Below this limit the index remains constant
$\alpha_1 = -1/3$.  This is due the fact that, below the limit, the
dominant part of the low frequency emission is generated by electrons
with $\gamma \simeq \gamma_{\rm brk}$.  Therefore, the spectrum in the
discussed range is equivalent to the spectrum produced by a
monoenergetic population of electrons, which can be approximated by a
power law with $\alpha=-1/3$ below the maximum of the emission
(e.g. \citealt{Rybicki79}). This means that even in the extreme 
case of the truncated, single power law energy distribution ($n_1=-\infty$)
there is always a ``source'' of low energy photons due to the low 
frequency emission ($F_{\rm syn} \sim \nu^{1/3}$) of the high energy 
electrons ($\gamma \simeq \gamma_{\rm brk}$). The spectral index of 
the inverse--Compton emission in the Thompson regime is equivalent 
to the index of the target radiation field (e.g. \citealt{Rybicki79}).
Since $\alpha_1$ is limited to --1/3 also the index of the 
inverse--Compton emission below the peak is limited to this value.
{\kk Moreover, this index does not depend on $n_2$.}
The index of the inverse--Compton emission above the peak is given by 
$\alpha=2 \alpha_2 - \alpha_1$, where $\alpha_1=-1/3$. This relationship 
can be easily derived from the approximation of the scattering derived by
\citet{Tavecchio98}.

\begin{figure}
\resizebox{\hsize}{!}{\includegraphics{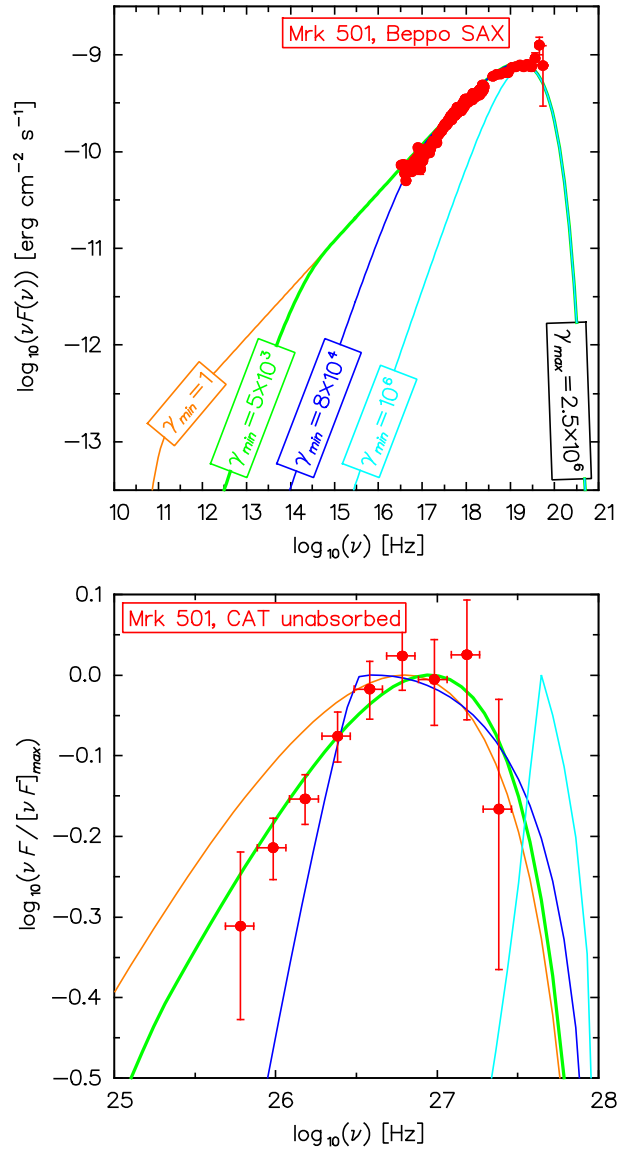}}
 \caption{Synchrotron emission (upper panel) and the self--Compton
          radiation (lower panel) for different values of the
          $\gamma_{\rm min}$ parameter. The best fit for the
          $Beppo$SAX (Pian et al. 1998) and CAT (Djannati-Atai et
          al. 1999) observations of Mrk 501 is indicated by the green
          lines. Note that the inverse--Compton spectra have been
          normalized and the TeV observations have been unabsorbed
          according to the model proposed by Kneiske et al. (2004).  }
\label{fig_gmt}         
\end{figure}

To directly check the applicability of the SSC scenario that assumes 
$n_1< 2$ we have applied our computations to the emission of 
1ES~1101--232 observed by $Beppo$SAX \citep{Wolter00} and
H.E.S.S. \citep{Aharonian06} experiments. The TeV data (here and in
all the other cases in the following) have been unabsorbed according
to the ``best-fit'' model proposed recently by \citet{Kneiske04}. The 
predictions of this model are also consistent with those of the model 
proposed by \citet{Stecker05}. The model provides a significantly stronger
absorption level than the minimal possible absorption suggested by
\citet{Aharonian06}. 
However, our best fit for the expected intrinsic spectrum, in a
framework of the above mentioned SSC scenario,  has been obtained for 
the limiting case ($n_1=-\infty$). This indicates that the absorption 
is close to the highest possible value that can be accommodated by 
this scenario. If the absorption is stronger, as suggested by some 
models (e.g. \citealt{DeJager02}), then our SSC scenario cannot explain 
the intrinsic TeV emission of 1ES~1101--232. The detailed values of 
the physical parameters used in our computations are discussed in 
the last section of this work, where we apply this scenario also to 
other sources.

The truncated particle energy distribution corresponds to the limit
$n_1 \rightarrow -\infty$ of a broken power law. 
For clarity, we will refer in the following to the break energy as
$\gamma_{\rm min}$. For such a electron distribution, the shape of the
TeV emission depends strongly on $\gamma_{\rm min}$, and we will
therefore investigate this is some detail.

\begin{figure*}
\resizebox{\hsize}{!}{\includegraphics{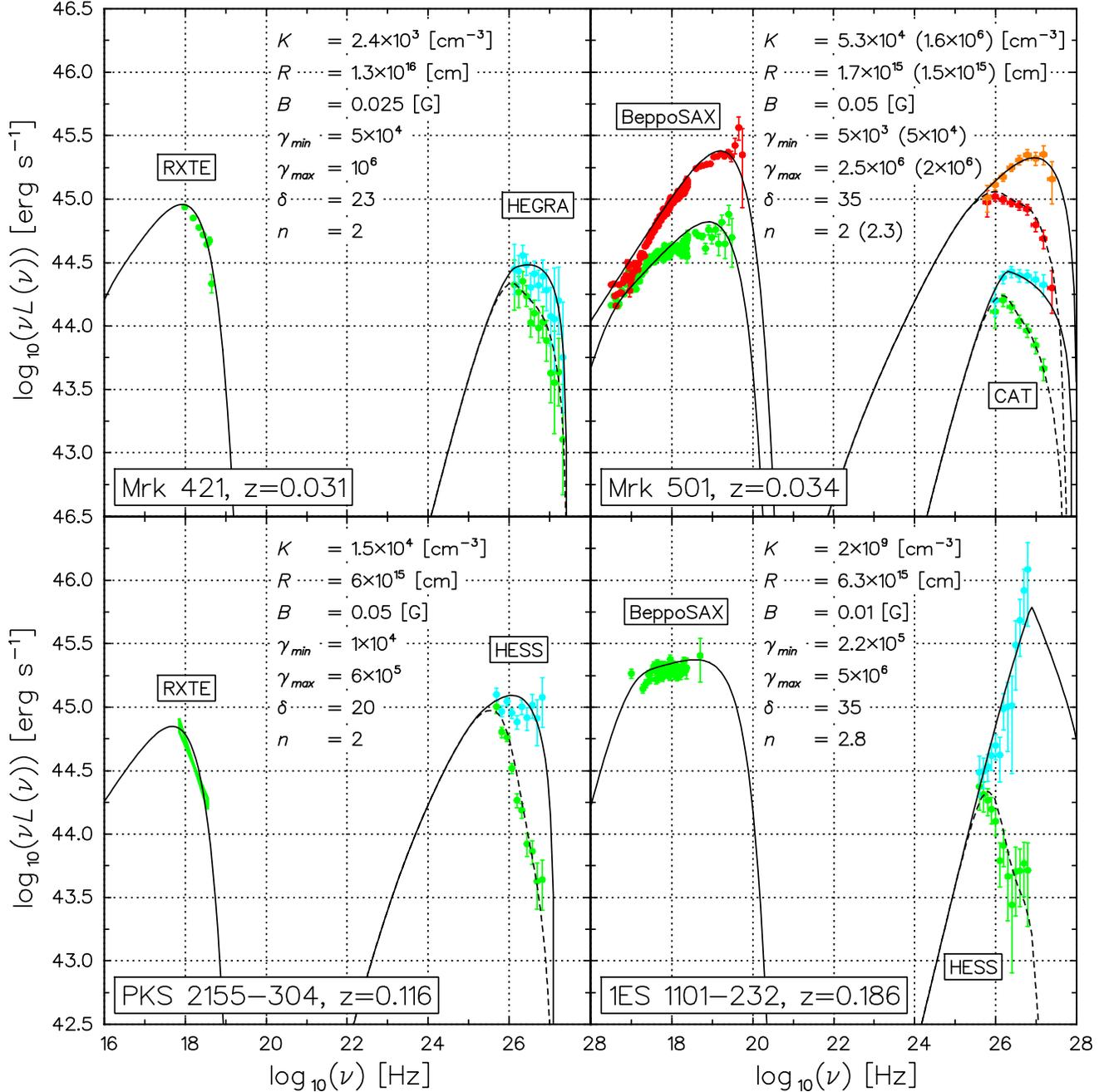}}
\caption{X--ray and TeV emission of four TeV blazars located at
         different redshifts. For all the sources we have applied the
         same SSC scenario, assuming that the emitting electrons
are distributed in energy as a single power law with a low energy
cut-off at $\gamma_{\rm min}$. All the TeV spectra have
         been unabsorbed according to the model of Kneiske et
         al. (2004). Solid lines are the best fits for the expected
         intrinsic emission whereas dashed lines indicate the observed
         spectra. For an easier comparison of the different sources, 
         we transformed the observed flux in luminosity.}
\label{fig_lum}         
\end{figure*}

\subsection{Changing $\gamma_{\rm min}$}

As we have shown in the previous subsection a particle distribution
with a low energy cut-off provides the best explanation for the
emission of 1ES~1101--232. We here show the effects of changing the
value of the low energy cut--off, which we will call
$\gamma_{\rm min}$.  
The result of our calculations is shown in
Fig. \ref{fig_gmt}. We have applied this test for the $Beppo$SAX
\citep{Pian98} and CAT \citep{Djannati99} observations of Mrk~501 from
16$^{\rm th}$ April, 1997. 

For this test, we fix the slope of the particle distribution to $n=2$,
which provides the best fit for the used X--ray observations of
Mrk~501.  In the first approach we assume the minimal possible value
of $\gamma_{\rm min}=1$.  This means that the model is equivalent to
the standard SSC scenario and is providing a relatively broad peak for
the inverse--Compton emission. Nevertheless, this approach is able to
roughly explain the unabsorbed TeV spectrum. However, the increase of
the $\gamma_{\rm min}$ value by about three orders of magnitude
provides a slightly narrower peak and therefore better explanation for
the spectrum.  A value of $\gamma_{\rm min}$ greater by one more order
of magnitude is the limiting value if we want to explain the entire
X--ray spectrum.  However, this already high value of $\gamma_{\rm
min}$ is drastically modifying the TeV spectrum, where the limiting
slope ($\alpha= -1/3$) of the emission becomes well visible (blue line
in Fig. 2).  Finally, a quasi--monoenergetic electron distribution
cannot explain the X--ray nor the TeV emission.  Note that for values
of $\gamma_{\rm max}$ close to $\gamma_{\rm min}$, the second part of
the TeV spectrum, above the peak, is curved and it is not a power law.

\section{Application to other TeV blazars}

We have already applied our specific SSC scenario to 1ES~1101--232 and
Mrk~501. In this section we apply this model to two other TeV sources,
Mrk~421 and PKS~2155--304, and compare the results.

The RXTE and HEGRA observations of Mrk~421 \citep{Krawczynski01} can
be fitted assuming that the X--ray emission is done by the exponential
synchrotron tail of the highest energy particles at $\gamma_{\rm
max}$.  Therefore, the index of the particle energy spectrum and
$\gamma_{\rm min}$ are not constrained by the data. The same approach
can explain the RXTE and H.E.S.S.  observations of PKS~2155--304
\citep{Aharonian05}.  Note that it is also possible to explain these
observations, as well as the observations of Mrk~501, assuming
$\gamma_{\rm min} \simeq 1$. We performed our computations to show
that the assumption of having $\gamma_{\rm min} \gg 1$ (required to
explain the emission for 1ES~1101--232), works very well also for
other sources. In Fig. \ref{fig_lum} we show the results of our
modeling and give the values of the input parameters. Note that for
Mrk~501 we provide also the spectral fits for the observations made on
April 7$^{\rm th}$, 1997. Moreover, in the case of Mrk~421, Mrk~501
and PKS~2155--304 the X-ray and TeV observations were made almost
simultaneously, whereas for 1ES~1101--232 we use the X--ray spectrum
from January 1997 and the TeV data collected over March-June
2004-2005.

{\kk 
Note that the value of the magnetic field used in our calculations is 
significantly smaller that the value estimated from the equipartition 
between the magnetic field energy density and the electron energy 
density. 
}
\section{Discussion}

We have shown that the standard SSC model can explain the specific
observations of the relatively distant TeV blazar
1ES~1101--232. Assuming a truncated power law energy distribution,
with $\gamma_{\rm min} \gg 1$, we can well explain the intrinsic TeV
spectrum even assuming a stronger absorption than the upper limit
suggested by \citet{Aharonian06}. This assumption appears in very
good agreement with the recent estimates of the IR intergalactic
background (\citealt{Kneiske04}, \citealt{Stecker05}).  The limit on
how hard an SSC spectrum can be is $\alpha=-1/3$.
This limit in turn translates in a limit on the amount of absorption 
suffered by TeV photons, and hence on the level of the IR background. 

Concerning the reason of having $\gamma_{\rm min} \gg 1$, 
we can propose two answers. The first is the injection
of an already truncated power law spectrum from a shock region.
If, in addition, the cooling process is inefficient for the 
the particles at lower energies (close to $\gamma_{\rm min}$),
the particle distribution will not develop a low energy tail
in a ``reasonable" time, namely a dynamical time,
over which we can assume that the physical conditions
(size, densities, magnetic field) are constant
(since the source has bulk motion, after a dynamical time
it will be located in a larger part of the jet,
and not contribute any longer to the observed flux).
{\kk
Note that in our particular simulations also particles 
with $\gamma \simeq \gamma_{\rm max}$ are not affected by
the cooling in one crossing time ($R/c$).
}

In a second, more complex, scenario all particles are accelerated 
efficiently at the shock front almost up to some maximal energy.  
Then they escape downstream of the shock and cool.  
However, the cooling process may be compensated (at least in part) 
by some stochastic turbulent acceleration \citep{Katarzynski05b}.  
The competition between re-acceleration (heating) and cooling may 
single out an equilibrium energy where the two processes balance.
This equilibrium energy may correspond to the $\gamma_{\rm min}$ 
assumed in our simple SSC scenario.

The changes of $\gamma_{\rm min}$ and $\gamma_{\rm max}$ may
have various impacts for the source variability, especially for 
the correlation between the X--ray and TeV emission. 
As an example, when
$n_2 < 3$ and $\gamma_{\rm min} \ll \gamma_{\rm max}$ the
change of $\gamma_{\rm min}$ may not affect the X--ray emission 
while having
a strong impact for the TeV radiation. 
Such a scenario could explain the orphan TeV flare in 1ES~1959+650 
\citep{Krawczynski04}. On the contrary, a change of $\gamma_{\rm max}$ 
would produce a significant X--ray variability without any change in the 
TeV flux below the TeV peak.

\section*{Acknowledgments}

We are grateful to A. Djannati-Atai, H. Krawczynski and B. Giebels for the data 
obtained by the CAT, HEGRA and H.E.S.S. experiments. We acknowledge the EC funding 
under contract HPRCN-CT-2002-00321 (ENIGMA network). JG thanks for the hospitality 
at the Osservatorio di Brera.

\label{lastpage}


\begin{thebibliography}{99}

\bibitem[\protect\citeauthoryear{Aharonian et al.}{2005}]{Aharonian05} 
Aharonian F., Akhperjanian A.G., Bazer--Bachi A.R., et al. 2005, A\&A, 442, 895

\bibitem[\protect\citeauthoryear{Aharonian et al.}{2006}]{Aharonian06} 
Aharonian F., Akhperjanian A.G., Bazer--Bachi A.R., et al. 2006, Nat, in press
                  (astro-ph/0508073)

\bibitem[\protect\citeauthoryear{Costamante et al.}{2003}]{Costamante03} 
Costamante L., Aharonian F.,  Ghisellini G. \& Horns D.      2003, New Astron. Rev., 
       47, 677

\bibitem[\protect\citeauthoryear{De Jager \& Stecker}{2002}]{DeJager02}
De Jager O.C., \& Stecker F.W.                           2002, 566, 738

\bibitem[\protect\citeauthoryear{Djannati-Atai et al.}{1999}]{Djannati99}
Djannati--Atai A., Piron F., Barrau A., et al.,            1999, A$\&$A, 350,  17 

\bibitem[\protect\citeauthoryear{Dwek \& Krennrich}{2005}]{Dwek05}
Dwek E. \& Krennrich F.,                                  2005, ApJ, 618,  657

\bibitem[\protect\citeauthoryear{Gould}{1966}]{Gould66}
Gould R.~J., Schreder G.~P.,                                 1966, Phys. Rev. Lett. 16, 252

\bibitem[\protect\citeauthoryear{Hauser \& Dwek}{2001}]{Hauser01}
Hauser M.~G., Dwek E., 2001, ARA\&A, 39, 249

\bibitem[\protect\citeauthoryear{Katarzynski et al.}{2005a}]{Katarzynski05a}
Katarzynski K., Ghisellini G., Tavecchio F., Maraschi L., Fossati G. \&
       Mastichiadis A.      2005a, A\&A, 433, 479

\bibitem[\protect\citeauthoryear{Katarzynski et al.}{2005b}]{Katarzynski05b}
Katarzynski K., Ghisellini G., Mastichiadis A., 
    Tavecchio F. \& Maraschi L.     2005b, A\&A, in press

\bibitem[\protect\citeauthoryear{Kneiske et al.}{2004}]{Kneiske04}
Kneiske T.M., Bretz T., Mannheim K., \& Hartmann D. H., 2004, A\&A, 413, 807


\bibitem[\protect\citeauthoryear{Krawczynski et al.}{2001}]{Krawczynski01}
Krawczynski H., Sambruna R., Kohnle A., et al.,           2001, ApJ, 559, 187

\bibitem[\protect\citeauthoryear{Krawczynski et al.}{2004}]{Krawczynski04}
Krawczynski H., Hughes S.B., Horan D., et al.            2004, ApJ, 601, 151

\bibitem[\protect\citeauthoryear{Madau \& Pozzetti}{2000}]{Madau00}
Madau P. \& Pozzetti L.,                                       2000, MNRAS, 312, L9 

\bibitem[\protect\citeauthoryear{Nishikov}{1962}]{Nishikov62}
Nishikov A.~I.,                                              1962, Sov. Phys. JETP, 14, 393

\bibitem[\protect\citeauthoryear{Pian et al.}{1998}]{Pian98}
Pian E., Vacanti G., Tagliaferri G., et al.,              1998, ApJ   , 492, L17

\bibitem[\protect\citeauthoryear{Rybicki \& Lightman}{1979}]{Rybicki79}
Rybicki G. \& Lightman A.P.,                             1979, Radiative Processes in Astrophysics, Wiley Interscience, New York

\bibitem[\protect\citeauthoryear{Stecker et al.}{1992}]{Stecker92}
Stecker F.~W., de Jager O.C. \& Salamon M.~H.,             1992, ApJ, 390, L49 

\bibitem[\protect\citeauthoryear{Stecker et al.}{2005}]{Stecker05}
Stecker F.~W., Malkan, M. A. \& Scully, S. T.,              2005, ApJ, submitted,
  (astro-ph/0510449)

\bibitem[\protect\citeauthoryear{Tavecchio et al.}{1998}]{Tavecchio98}
Tavecchio F., Maraschi L. \& Ghisellini G.,             1998, ApJ   , 509, 608

\bibitem[\protect\citeauthoryear{Wolter et al.}{2000}]{Wolter00}
Wolter A., Tavecchio F., Caccianiga A. \& Tagliaferri G.,           2000, A\&A, 357, 429

\end{thebibliography}
\end{document}